\documentclass[%
 reprint,
showpacs,
showkeys,
 amsmath,amssymb,
 aps,
]{revtex4-1}

\usepackage{graphicx}
\usepackage{dcolumn}
\usepackage{bm}

\begin{document}

\title{Asymmetry of radiation damage properties in Al-Ti nanolayers}

\author{Wahyu Setyawan}
 \email{corresponding author: wahyu.setyawan@pnnl.gov}
\author{Matthew Gerboth}
\author{Bo Yao}
\author{Charles H. Henager, Jr.}
\affiliation{Pacific Northwest National Laboratory, Richland WA 99354, U.S.A.}
\author{Arun Devaraj}
\author{Venkata R. S. R. Vemuri}
\author{Suntharampillai Thevuthasan}
\author{Vaithiyalingam Shutthanandan}
\affiliation{Environmental Molecular Sciences Laboratory, Richland WA 99354, U.S.A.}

\date{\today}
\begin{abstract}
Molecular dynamics (MD) simulations were employed with empirical potentials to study the effects of multilayer interfaces and interface spacing in Al-Ti nanolayers. Several model interfaces derived from stacking of close-packed layers or face-centered cubic \{100\} layers were investigated. The simulations reveal significant and important asymmetries in defect production with $\sim$60\% of vacancies created in Al layers compared to Ti layers within the Al-Ti multilayer system. The asymmetry in the creation of interstitials is even more pronounced. The asymmetries cause an imbalance in the ratio of vacancies and interstitials in films of dissimilar materials leading to $>$90\% of the surviving interstitials located in the Al layers. While in the close-packed nanolayers the interstitials migrate to the atomic layers adjacent to the interface of the Al layers, in the \{100\} nanolayers the interstitials migrate to the center of the Al layers and away from the interfaces. The degree of asymmetry and defect ratio imbalance increases as the layer spacing decreases in the multilayer films. Underlying physical processes are discussed including the interfacial strain fields and the individual elemental layer stopping power in nanolayered systems. In addition, experimental work was performed on low-dose (10$^{16}$ atoms/cm$^2$) helium (He) irradiation on Al/Ti nanolayers (5 nm per film), resulting in He bubble formation $\sim$1 nm in diameter in the Ti film near the interface. The correlation between the preferential flux of displaced atoms from Ti films to Al films during the defect production that is revealed in the simulations and the morphology and location of He bubbles from the experiments is discussed.
\end{abstract}

\keywords{radiation damage asymmetry, self-healing nanolayer, Al-Ti interatomic potential}
\maketitle

\section{Introduction}
Radiation damage in solids from collision cascades formed during high-energy particle irradiation, ions or neutrons, is extremely costly and, perhaps, the single most complex and challenging technological problem facing nuclear material scientists, reactor designers, and regulatory agencies desiring long-lived engineering structures, low operational costs, and safety. The search for and development of materials with improved radiation damage tolerance requires a more or less complete understanding of defect production, transport, evolution, and recovery in complex alloy or composite systems that are undergoing irradiation and transitions far from equilibrium on picosecond time scales at the atomic level to decade-long microstructural and thermo-physical property changes as either structural or functional materials. These properties undergo unavoidable time-, temperature-, and fluence-dependent degradation and, usually, irreversible changes \cite{WasBook} such that replacement or costly mitigation is required to satisfy operational safety concerns at critical fluence levels. There is a strong scientific and technological interest in studying and developing radiation damage tolerant materials.

Structural materials can achieve radiation damage tolerance via two basic mechanisms. Some materials intrinsically have a damage tolerant crystal structure with high damage thresholds, such as SiC in the zinc-blende structure, or they possess a high tolerance for atomic disorder as evidenced by certain oxides, such as disordered fluorites \cite{Sickafus2007NatMat}. Unfortunately, most metallic and structural alloys possess low damage tolerance from close-packed crystal structures that can accommodate a wide variety of low lying defect states and that have low damage thresholds. Thus, they do not possess intrinsic damage tolerance, although bcc materials are more damage tolerant compared to fcc or hcp structures \cite{WasBook}.

The second basic mechanism relies on enhanced damage recovery mechanisms typically via increased recombination rates of radiation-induced defects at defect sinks within a material. These sink sites range from grain boundary denuded zones observed in many materials to engineered materials containing nano-spaced interfaces, including nano-featured ferritic alloys and nanolayered materials. The general concept of point defect recombination at internal interfaces is not a new idea but has achieved recent significance from work with oxide dispersion strengthened (ODS) alloys \cite{Lescoat2012JNM, He2012JNM, Hsiung2011JNM, Miller2006JNM, Ukai2007EnergyMat, Lindau2005FusionEng}, nano-featured alloys \cite{Odette2008ARMR, Wu2012ActaMat}, and nano-layered composites \cite{Heinisch2004JNM, Misra2007JOM, Zhang2012ASME} specifically designed to achieve high strength and enhanced defect recombination at closely spaced sinks for vacancies and self-interstitials. Capture and immobilization of helium (He) is also of keen interest for fusion reactor materials where He can be produced at levels approaching a few atomic percent \cite{Edmondson2011ScriptaMat, Odette2010JOM, Kashinath2013PRL, Demkowicz2012CurrentOp, Zhernenkov2011APL} .

Specifically, nanolayered materials based on dissimilar materials arranged in closely spaced layered structures with high interfacial areal fractions are considered developmental radiation tolerant materials. The specific details of the damage tolerance are still being studied and evaluated but it is considered that enhanced defect recombination at the dissimilar interfaces is occurring that reduces the overall damage accumulation relative to bulk materials. However, some specific trends are noted and discussed in the literature, namely, that immiscible systems behave differently compared to miscible systems under irradiation \cite{Fu2013PhilMag}. Miscible systems, including Al-Ti reported here, intermix under irradiation and would not be expected to demonstrate radiation damage tolerance at high doses. Immiscible systems are stable against mixing and do demonstrate enhanced radiation damage tolerance to some level of damage \cite{Fu2013PhilMag}. With regard to mixing of layered materials, but not specifically nanolayered films, there are phenomenological models built on the assumption that the mixing occurs via interdiffusion during high-energy collision cascades at low temperatures where the ion beam supplies sufficient energy that a locally melted region develops (thermal spike region) and phase transitions are possible \cite{Cheng1990MatSciRep}. These models and the thermodynamics of mixing can partly explain the improved damage tolerance of the immiscible systems compared to miscible ones.

However, many collisions are lower in energy than considered for the interdiffusion mixing models and in this regime a systematic study of nanolayered materials and their response to radiation damage has been lacking. At lower energies we can partly avoid the complications of ion beam mixing and study more carefully the effects of displacement damage. In this respect we find that there has been a lack of theoretical studies in this regime and this paper focuses on this aspect of the problem for a specific layered system that can be arranged in atomic models in a wide variety of stable structures, namely, the Al-Ti system. We also include some preliminary He-ion implantation studies of sputtered nanolayered Al-Ti films that demonstrate agreement with the theoretical models studied here.

The use of low energy He ion implantation to study ion beam mixing and radiation damage in nanolayered thin films is useful since He damage rates are reduced compared to heavy ions, the ranges are appropriate for thin films, and the effects of He accumulation are relatively easily observed compared to point defect clustering as a measure of radiation damage. It is understood that He bubble formation proceeds from vacancy (V) accumulation and He-V binding. Thus, observing He bubbles is a surrogate for observing V clustering in these thin film materials. H\"{o}chbauer et al. \cite{Hochbauer2005JAP} were the first to study He accumulation as bubbles in Cu-Nb nanolayered materials. They observed preferential He bubble formation at Cu-Nb interfaces and along columnar grain boundaries following 33 keV He implantation. Demkowicz et al. \cite{Demkowicz2007NIMB} concluded that He also accumulates along Cu-Nb interfaces and that these interfaces act as fast diffusing pathways for He escape during annealing.

Zhang et al. \cite{Zhang2007NIMB} observed that He bubbles were not resolvable in Cu-Nb 2.5-nm layered foils, whereas identical 33 keV He implantation produced TEM visible bubbles in pure Cu, pure Nb, and Cu-Nb 100-nm layered materials. This was assumed to be evidence that Cu-Nb 2.5-nm layered materials exhibited enhanced recombination of radiation-induced point defects and, thus, much smaller He bubbles, less than about 1 nm in diameter. Zhernenkov et al. used neutron reflectometry to study He locations in implanted Cu-Nb foils and concluded that He was likely being stored as interstitial He in the dissimilar interfaces until a critical concentration was reached, after which He bubbles were formed \cite{Zhernenkov2011APL}. Perhaps the best evidence comes from $^3$He implantations and using nuclear reaction analysis (NRA) to study He concentrations as a function of implantation depth together with TEM to determine He concentrations where He bubbles form \cite{Demkowicz2010APL}. Similar conclusions were reached by Bhattacharyya et al. using TEM and NRA to study $^3$He-implanted Cu-Nb foils \cite{Bhattacharyya2012MM}. Interface structure appears to play a critical role in the amount of He that can be stored before bubbles form \cite{Demkowicz2012CurrentOp}. Recent MD studies are consistent with this understanding and demonstrate atomic storage mechanisms for He in certain interfaces \cite{McPhie2013JNM, Kashinath2013PRL}.

However, once a critical concentration of He is reached then He bubbles can nucleate and grow in these layered materials just as in bulk metals. A key difference, though, is that He bubble morphologies and locations vary from layer to layer and, above a certain dose, appear to depend on some intrinsic property of the layer material rather than the interfaces \cite{Li2012Plasticity}. Hattar et al. \cite{Hattar2008Scripta} observed He bubbles in both the Cu and Nb layers of a Cu-Nb 5 to 6-nm layered foil after high doses of 33 keV He ion implantation at 763 K. However, He bubbles in the Cu layers spanned the thickness of the entire layer and were approximately 5 to 6 nm in diameter, whereas He bubbles in the Nb layers were about 1 to 2 nm in diameter. Similar observations of He bubble suppression compared to bulk or 100-nm layered materials are observed in Cu-V nanolayered foils \cite{Fu2009JNM} and in Cu-Mo nanolayered foils \cite{Li2011PhilMagLett}, where a slight size difference between He bubbles in Cu layers (larger) compared to Mo layers was noted. Wei et al. \cite{Wei2011PhilMag} observed bubble size differences in Ag-V nanolayered materials somewhere between the Cu-Nb size differences and those observed for Cu-V, with the larger bubbles contained in Ag layers. Fu et al. nicely summarize dose effects in Cu-V nanolayered systems and discuss He effects, radiation hardening, and both mixing and demixing effects observed in other systems \cite{Fu2013PhilMag}.

One trend that appears to be consistent in these nanolayered studies is the observation that a certain level of asymmetry develops with regard to He bubble morphologies at increased He doses. Bubble sizes are non-uniform after a certain dose and the evidence is not clear that this asymmetry does not develop earlier in the radiation damage regime. Helium storage at dissimilar interfaces does not destroy the symmetry of the system, although, asymmetric swelling amounts are often noted \cite{Zhernenkov2011APL, Li2012Plasticity, Hattar2008Scripta, Fu2009JNM}, along with asymmetric He bubble sizes \cite{Hattar2008Scripta}. These become serious issues in dealing with nanolayered failure mechanisms from radiation damage, perhaps from delamination or other mechanical failures due to differential responses.

One shortcoming in the current literature and that is addressed in this research is the lack of understanding of point defects in nanolayered systems at low energies where ion beam mixing and demixing effects do not occur readily. In particular, displacement thresholds have not been studied for any of these layered systems to help understand or predict if some part of the response asymmetry may be due to displacement threshold effects. There is no reason to expect that defect generation or fates are symmetric within nanolayered materials made up of dissimilar metals. Under asymmetric defect generation the ability of the system to avoid damage accumulation via enhanced recombination may be compromised. One layer may accumulate an excess of one kind of point defect or defect cluster over time. The differential He bubble size observed in Cu-Nb suggests that this type of damage cannot be overlooked or ignored. An analogy to the Kirkendall effect and the resultant porosity and interface motion during interdiffusion of binary diffusion couples may be helpful.

This study performs a series of MD simulations in the Al-Ti system, which is one that has not been studied in terms of radiation damage response. The choice of this system was motivated by the availability of a high-quality EAM potentials and by the flexibility of this system, although it is extremely reactive and miscible, to adopt a variety of possible interfaces, namely fcc-fcc at small size scales and fcc-hcp at larger size scales. In addition, this system is readily synthesized using magnetron sputtering. In a separate publication we set out a method and data for displacement thresholds for the Al-Ti nanolayered system and see systematic differences between computed thresholds for bulk metals and nanolayered metals. We make use of this information here but the research reported here studies the radiation response of the Al-Ti system arranged in a variety of possible structures. Finally, we have some preliminary data on He-implanted Al-Ti layers using low energy He ions using a He ion microscope and with characterization in cross-section using FIB and TEM.

\section{Methods}
\subsection{Computational details}
\subsubsection{Interatomic potentials: modification}

As mentioned above, one motivation in choosing Al-Ti systems was the recent availability of the high quality Al-Ti embedded-atom (EAM) potentials developed by Zope {\it et al} \cite{Zope2003PRB}. The potentials were fitted to a large database of experimental as well as ab initio data. A comprehensive list of properties was reproduced accurately. Those that are particularly important for radiation damage in multilayers include the vacancy formation and migration energies, elastic moduli, stacking fault energies, and the formation energy of various bulk phases. In addition, the potentials yielded accurate coefficients of thermal expansion. Hence, isobaric-isothermal ({\it NPT}) simulations can be performed. In order to use the potentials for simulating radiation damage, the short-range parts need modification to accurately model the highly repulsive interactions that dominate the early stages of collision cascades. The modification was applied to the pair interactions of the EAM potentials.

\begin{figure}
\includegraphics[scale=0.85]{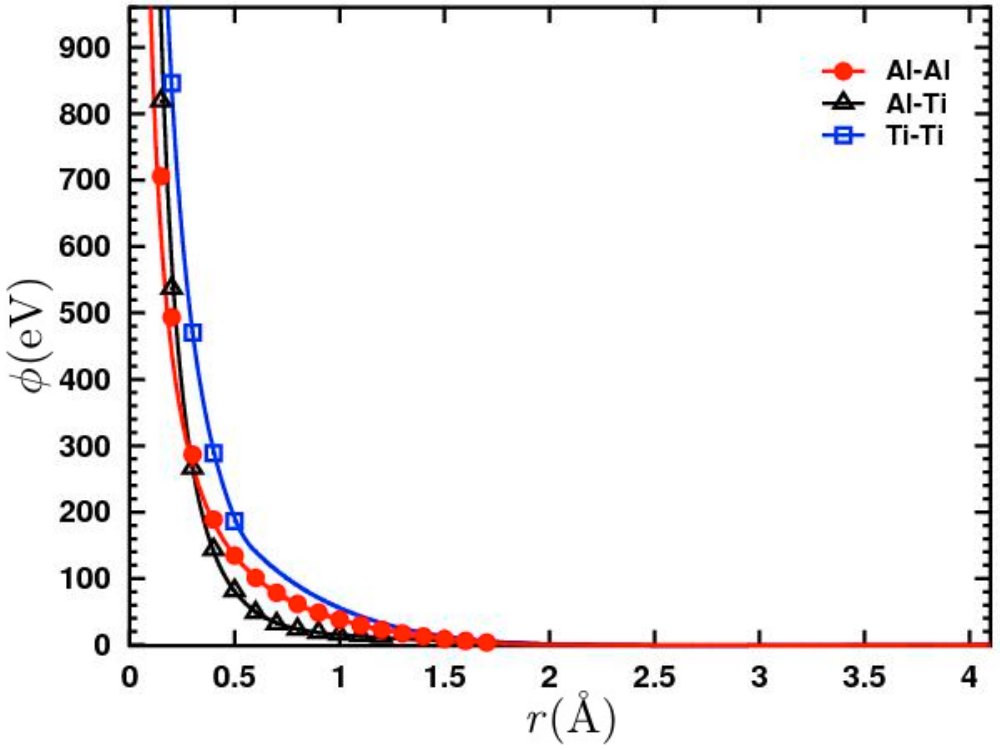}
\caption{\label{fig_modzope} Short-range part of the pair interactions modified from the original Al-Ti embedded-atom potentials \cite{Zope2003PRB}. The data points denote the ab initio energies used for fitting.}
\end{figure}

For the short-range modification, ab initio energies of dimers Al-Al, Al-Ti and Ti-Ti at various bond lengths were calculated and used for fitting. VASP software was utilized to perform the first-principles calculations within the density-functional-theory (DFT) formalism using plane-wave bases \cite{vasp1,vasp2}. Accurate projector-augmented-wave pseudopotentials with Perdew-Burke-Ernzerhof exchange correlations were used \cite{PAW,PAWkresse,PBE}. The plane-wave energy cutoffs were 240.30 and 178.33 eV for Al and Ti respectively. The number of electrons treated as valence electrons was three for Al and four for Ti. To simulate an isolated two-body system, {\bf $\Gamma$}-point calculations were performed in a cubic box of side 15 \AA. With this setup, the interactions between periodic images were verified to be negligible. The self-consistent loop was converged with a tolerance of 0.1 meV. To extract the two-body interaction potentials, the appropriate atomic energies were subtracted from the total energy. With these modifications, there were three regions of pair interactions based on the distance between two atoms $r$:

\begin{eqnarray}
\phi & = & \left\{
\begin{array}{ll}
\phi_{mod}, & r \le r_{mod}\\
\phi_{sp},  & r_{mod} < r < r_{sp}\\
\phi_{zope}, &  r_{sp} \le r\\
\end{array}
\right.
\label{eq1}
\\
\phi_{mod} & = & \frac{Z_1 Z_2}{r}
\alpha e^{-\beta r/a} + \delta, 
a = \frac{0.4683}{Z_1^{0.23} + Z_2^{0.23}}
\label{eq2}
\\
\phi_{sp} & = & c_0 + c_1 \lambda + c_2 \lambda^2 + c_3 \lambda^3, \lambda = r - r_{mod}
\label{eq3}
\end{eqnarray}
In the above expressions, $\phi_{zope}$ denotes the original pair interaction, $\phi_{mod}$ denotes the part of pair interaction that is fitted to the ab initio data, and $\phi_{sp}$ represents a natural cubic spline interpolating $\phi_{zope}$ and $\phi_{mod}$ with $Z$ the atomic number. The nonlinear-least-square fitted values of $r_{mod}$, $r_{sp}$, $\alpha$, $\beta$ and $\delta$ as well as the coefficients of the splines are presented in Table \ref{table1}. The functional used in $\phi_{mod}$ (Eq. \ref{eq2}) follows the usual Ziegler-Biersack-Littmark (ZBL) parameterization \cite{ZBLbook,Hossain20091061}. Figure \ref{fig_modzope} shows the short-range part of the modified pair interactions along with the ab initio data points.

\begin{table*}
\caption{\label{table1}
Fitted parameters for the short-range part of the pair interactions modified from the original Al-Ti embedded-atom potentials \cite{Zope2003PRB}.}
\begin{ruledtabular}
\begin{tabular}{cccccc cccc}
 &$r_{mod}$ (\AA)&$r_{sp}$ (\AA)&$\alpha$(eV$\cdot$ \AA)&$\beta$(\AA$^{-1}$)&$\delta$(eV)& $c_0$(eV) & $c_1$(eV$\cdot$ \AA$^{-1}$) & $c_2$(eV$\cdot$ \AA$^{-2}$) & $c_3$(eV$\cdot$ \AA$^{-3}$) \\
\hline
Al-Al & 1.65096 & 1.97712 & 0.617098 & 0.082469 & -17.271753 & 4.85758 & -27.43908 & 57.31412 & -39.24883\\
Al-Ti & 1.00066 & 2.62203 & 0.868468 & 0.476070 & 10.274221 & 15.26817 & -24.51782 & 13.02075 & -2.41210\\
Ti-Ti & 0.50706 & 0.60457 & 0.485486 & 0.167387 & -35.530090 & 185.50781 & -755.59674 & 3877.36020 & -10709.64591\\
\end{tabular}
\end{ruledtabular}
\end{table*}

\subsubsection{Multilayer construction}
Five multilayer families (systems) were investigated: Mfcc, Mhcp, M100, Mcp and Mcpic. Within each system, four multilayers were constructed with different film thickness: three, six, 12 and 24 atomic layers per film. In this study, the keyword film refers to Al film or Ti film. The multilayer systems are designated as the following.
MfccL3 represents stacking of face-centered-cubic (fcc) \{111\} close-packed layers of Al and Ti with three layers per film. 
MhcpL6 represents stacking of hexagonal close-packed (hcp) \{0001\} layers of Al and Ti with six layers per film. 
M100L12 represents stacking of fcc \{100\} layers of Al and Ti with 12 layers per film. 
McpL24 represents stacking of fcc close-packed layers of Al and hcp close-packed layers of Ti with 24 layers per film. 
Mfcc, Mhcp, M100 and Mcp are multilayers with coherent interfaces. Mcpic multilayers are similar to Mcp only with incommensurate interfaces. Note that in an experimental work \cite{Saleh1997TionAl001}, fcc Ti grows epitaxially on Al(100) up to six layers , beyond which the axial alignment with the substrate is only partially preserved and off-normal alignment is lost, however the exact structure is unknown. On Al(111), Ti was experimentally determined to form a two-dimensional hcp overlayer up to two monolayers with an incommensurate interface, followed by three-dimensional island growth \cite{Kim1999}.

For the Mcp systems, the ground state stacking was determined via energy minimization with the conjugate-gradient technique as implemented in LAMMPS software \cite{LAMMPS}. The repeat unit was found to be abcABAbacBAB for the McpL3 and abcabcABABAB for the McpL6 (a lower or upper case denotes Al or Ti layer, respectively). The McpL12 and McpL24 multilayers are simply extensions of the McpL6. The stacking in Mcpic system follows that in Mcp. For comparison, four bulk structures were constructed: fccAl, hcpAl, fccTi and hcpTi. The lattice vectors of the simulation cells are denoted as {\bf L}$_1$ and {\bf L}$_2$ spanning the basal dimensions and {\bf L}$_3$ along the stacking direction. For all the coherent multilayers, cubic or nearly-cubic orthorhombic cells were used containing 55,296 atoms arranged in 48 layers, with {\bf L}$_1$, {\bf L}$_2$ and {\bf L}$_3$ along $x$, $y$, and $z$ axes, respectively. The Mcpic systems were generated as follows. Starting from the ground state fcc Al and hcp Ti, the Al film was constructed using lattice vectors {\bf a$_1$} = [2.8636, 0, 0], {\bf a$_2$} = 2.8636$\times$[$\frac{1}{2}$, $\frac{1}{2}\sqrt{3}$, 0] and {\bf a$_3$} = [0, 0, 2.3386] while the Ti film was constructed using lattice vectors {\bf b$_1$} = [2.9529, 0, 0], {\bf b$_2$} = 2.9529$\times$[$\frac{1}{2}$, $\frac{1}{2}\sqrt{3}$, 0] and {\bf b$_3$} = [0, 0, 2.3402]. Each Al layer was generated from 34$\times$33 supercell (1122 atoms per atomic layer), while 33$\times$32 supercell was done for Ti (1056 atoms per atomic layer). The total number of atoms in the Mcpic system was 52,272 (48 layers). The basal dimensions for the simulation cell were taken from the Al supercell, i.e. {\bf L}$_1$=34$\times{\bf a}_1$ and {\bf L}$_2$=33$\times{\bf a}_2$. Note that, even though the misfit was greatly minimized by such supercell sizes, it cannot be eliminated due to the incommensurability. Unlike in the coherent multilayers in which the Al film is exclusively under tensile (while Ti is exclusively compressed) in all basal directions, the Mcpic so constructed with an unequal length of basal vectors was thought to minimize such an exclusive strain in a particular film and to better model an unconstrained incommensurate system. The dimensions of all the structures and the strains in each layer at 300 K are presented in Table \ref{tablestrain}.

\begin{table*}
\caption{\label{tablestrain}
Dimensions of Al-Ti multilayers and bulk Al and Ti structures at 300 K. $V_{vor}$ represents the average Voronoi volume per atom. $\bar{d}_z$ denotes the average Voronoi thickness of a layer in each film. The strains are calculated relative to the constituent bulk structure of Al or Ti in the multilayers (e.~g. in Mfcc system the bulk structures are fccAl and fccTi while in Mcp system they are fccAl and hcpTi). $\Delta V = V_{vor} - V$, where $V$ denotes the average atomic volume in the constituent bulk structures.}
\begin{ruledtabular}
\begin{tabular}{c ccccc cccccc}
 &$L_x$ (\AA)&$\bar{d}_z^{Al}$ (\AA)&$\bar{d}_z^{Ti}$ (\AA)&$V_{vor}^{Al}$ (\AA$^3$)&$V_{vor}^{Ti}$ (\AA$^3$)
 &$\epsilon_x^{Al}$(\%)&$\epsilon_x^{Ti}$(\%)&$\bar{\epsilon}_z^{Al}$(\%)&$\bar{\epsilon}_z^{Ti}$(\%)&$\frac{\Delta V}{V}^{Al}$(\%)&$\frac{\Delta V}{V}^{Ti}$(\%) \\
\hline
M100L3 &24$\times$4.218&1.825&1.983  &  16.244&17.645  &  3.65&1.52  &  -10.25&-4.55  &  -3.60&-1.60 \\
M100L6 &24$\times$4.142&1.950&2.061  &  16.730&17.686  &  1.79&-0.30  &  -4.19&-0.80  &  -0.71&-1.38 \\
M100L12 &24$\times$4.129&1.986&2.078  &  16.925&17.707  &  1.45&-0.63  &  -2.40&0.01  &  0.45&-1.26 \\
M100L24 &24$\times$4.120&2.053&2.102  &  17.430&17.844  &  1.25&-0.83  &  0.91&1.18  &  3.45&-0.49 \\
\hline
MfccL3 &32$\times$2.906&2.287&2.376  &  16.742&17.393  &  1.00&-1.08  &  -2.65&-0.94  &  -0.64&-3.01 \\
MfccL6 &32$\times$2.913&2.304&2.393  &  16.933&17.587  &  1.23&-0.85  &  -1.95&-0.26  &  0.49&-1.93 \\
MfccL12 &32$\times$2.914&2.323&2.395  &  17.084&17.613  &  1.26&-0.82  &  -1.14&-0.18  &  1.39&-1.78 \\
MfccL24 &32$\times$2.919&2.279&2.425  &  16.829&17.900  &  1.45&-0.64  &  -2.98&1.07  &  -0.12&-0.18 \\
\hline
MhcpL3 &32$\times$2.896&2.359&2.398  &  17.144&17.422  &  1.68&-1.98  &  -3.91&1.81  &  -0.66&-2.16 \\
MhcpL6 &32$\times$2.904&2.388&2.390  &  17.445&17463  &  1.97&-1.70  &  -2.75&1.51  &  -1.08&-1.93 \\
MhcpL12 &32$\times$2.909&2.403&2.382  &  17.612&17.458  &  2.14&-1.54  &  -2.12&1.16  &  2.05&-1.96 \\
MhcpL24 &32$\times$2.918&2.343&2.387  &  17.281&17.605  &  2.46&-1.22  &  -4.57&1.36  &  0.13&-1.13 \\
\hline
McpL3 &32$\times$2.906&2.304&2.375  &  16.841&17.362  &  0.97&-1.66  &  -1.95&0.85  &  -0.05&-2.49 \\
McpL6 &32$\times$2.917&2.305&2.375  &  16.983&17.496  &  1.35&-1.29  &  -1.89&0.84  &  0.79&-1.74 \\
McpL12 &32$\times$2.922&2.318&2.368  &  17.136&17.508  &  1.54&-1.11  &  -1.33&0.57  &  1.70&-1.68 \\
McpL24 &32$\times$2.925&2.294&2.386  &  16.999&17.682  &  1.64&-1.01  &  -2.37&1.32  &  0.88&-0.70 \\
\end{tabular}
\begin{tabular}{cc ccccc cccccccc}
Mcpic &$L_1$&$L_2$&$\bar{d}_z^{Al}$&$\bar{d}_z^{Ti}$&$V_{vor}^{Al}$ &$V_{vor}^{Ti}$ 
 &$\epsilon_1^{Al}$(\%)&$\epsilon_1^{Ti}$(\%)&$\epsilon_2^{Al}$(\%)&$\epsilon_2^{Ti}$(\%)&$\bar{\epsilon}_z^{Al}$(\%)&$\bar{\epsilon}_z^{Ti}$(\%)&$\frac{\Delta V}{V}^{Al}$(\%)&$\frac{\Delta V}{V}^{Ti}$(\%) \\
 \hline
L3 &97.055&94.608 &2.332&2.370  &  16.455&17.770  &  -0.80&-0.46  &-0.36&0.05 &  -0.74&0.65  &  -2.34&-0.20 \\
L6 &97.332&94.666 &2.347&2.362  &  16.644&17.801  &  -0.52&-0.18  &-0.30&0.11 & -0.13&0.30  &  -1.22&-0.03 \\
L12 &97.473&94.710 &2.355&2.357  &  16.753&17.820  &  -0.37&-0.03  &-0.25&0.16 &  0.21&0.10  &  -0.57&-0.08 \\
L24 &97.510&94.717 &2.355&2.356  &  16.772&17.835  &  -0.34&0.01  &-0.24&0.17 & 0.21&0.06  &  -0.46&0.16 \\
\hline
 &fccAl&fccTi&hcpAl&hcpTi& & &&&&& \\
 \hline
$a$(\AA) &4.069&4.155&2.848&2.955& & &&&&& \\
$c/a$ &1.000&1.000&1.724&1.594& & &&&&& \\
\end{tabular}
\end{ruledtabular}
\end{table*}

\subsubsection{Molecular Dynamics simulations setup}
LAMMPS software was used to perform the MD simulations employing periodic boundary conditions (PBCs) in all dimensions. Before a displacement cascade was initiated, each structure was thermalized at 300 K and zero pressure ({\it NPT}) for 30 ps. To obtain a proper canonical distribution of velocity, the thermalization was performed using Nos\'{e}-Hoover thermostat with a time step of 0.5 fs and a 1-ps damping parameter \cite{Nose,Hoover}. To initiate a collision cascade, a random primary-knock-on atom (PKA) was chosen and was assigned an initial velocity normal to the stacking direction. Throughout this study, the PKA was given an initial kinetic energy of 1.5 keV. This PKA energy is sufficient to cause damage across most of the interfaces in the constructed multilayers and yet small enough to avoid overlaps of damage regions due to PBCs. The displacement cascade and damage recovery processes were simulated in five stages:
\vspace{-7pt}
\begin{enumerate}
\item  {\it Early collision} (0.025 ps): fix fixnve all nve; reset\_timestep 0; timestep 0.005E-3; run 5000.
\vspace{-9pt}
\item  {\it Creation of thermal-spike regions} (1 ps): timestep 0.02E-3; run 50000.
\vspace{-9pt}
\item {\it Cooling of thermal-spike regions} (0.5 ps): timestep 0.05E-3; run 10000.
\vspace{-9pt}
\item {\it Main recovery} (4 ps): timestep 0.2E-3; run 20000.
\vspace{-9pt}
\item {\it Migration and final thermalization} (50 ps): unfix fixnve; fix fixnvt all nvt temp 300.0 300.0 1.0; timestep 0.5E-3; run 100000.
\end{enumerate}
\vspace{-7pt}
Stages 1$\rightarrow$4 were performed in a constant-energy ($NVE$) condition. In stage 1, the timestep was so chosen that no atom moved beyond approximately 0.005 \AA\ per time step. Throughout the simulation, the temperature of the system was below 509 K and at the end of stage 2 the temperature was typically 390 K. During the last stage, the temperature was thermalized to 300 K with a damping factor of 1 ps.

The total simulation time was approximately 55.5 ps. For each structure, 20 runs were performed. In multilayer structures, ten runs with an Al PKA and ten runs with a Ti PKA were done. Within each system, only one initial thermalization run was performed. All damage cascades in this system were started from the same thermalization restart file. For defect counting analysis, a reference configuration was generated with molecular static energy minimization in each system. Voronoi cells were then constructed using these reference sites. Unoccupied cells were identified as vacancies and the number of vacancies was taken as the number of Frenkel pairs.

\subsection{Experimental techniques}
An Al/Ti multilayer thin film stack for a total thickness of 400 nm with individual layer thickness of 5 nm was fabricated on a cleaned silicon (100) substrate using direct current magnetron sputter deposition. The  base pressure of  the sputter deposition system was 5$\times$10$^{-8}$ Torr.  Individual layers of Ti and Al were deposited at cathode powers of 180 and 240 watt, respectively, with 2 mTorr argon process gas pressure. Helium implantation (30 kV) to a dose of 10$^{16}$ ions/cm$^2$ was performed using a He ion microscope on an area of 10$\times$10 $\mu$m$^2$. The total thickness of the stack was chosen in such a way that maximum damage is located in the center of the stack. The damage profile and maximum ion range were estimated using Stopping Range of Ions in Matter (SRIM) simulation (shown in Figure \ref{figHe}c) \cite{ZBLbook, BiersackNIM1980, srimweb}. For a He ion fluence of 10$^{16}$ ions/cm$^2$, the estimated peak damage was 0.375 dpa, which is located approximately at a depth of 180-nm from the surface.

After He implantation, a cross sectional transmission electron microscopy (TEM) lamella sample was fabricated using site specific FIB lift-out process. TEM imaging was performed using a JEOL 2010F TEM. Overfocused and underfocused TEM imaging was performed to image the He bubbles.  Helium bubbles show bright contrast in underfocused TEM images and darker contrast in overfocused images. The TEM images of the region between the top surface and peak helium implantation dose are shown in Figure \ref{figHe}. In the underfocused image, the bubbles can be clearly seen to be preferentially segregated to the darker contrast Ti layers. Furthermore, a spatial distribution of He bubbles within the Ti layer closer to the interface of Ti/Al is also observed. Surprisingly the Al layer did not appear to have any bubbles or the bubble size is below the TEM resolution, which is estmated to be about 1-nm.

\section{Results and Discussion}
\subsection{Simulation Results}
\begin{figure}
\includegraphics[scale=0.75]{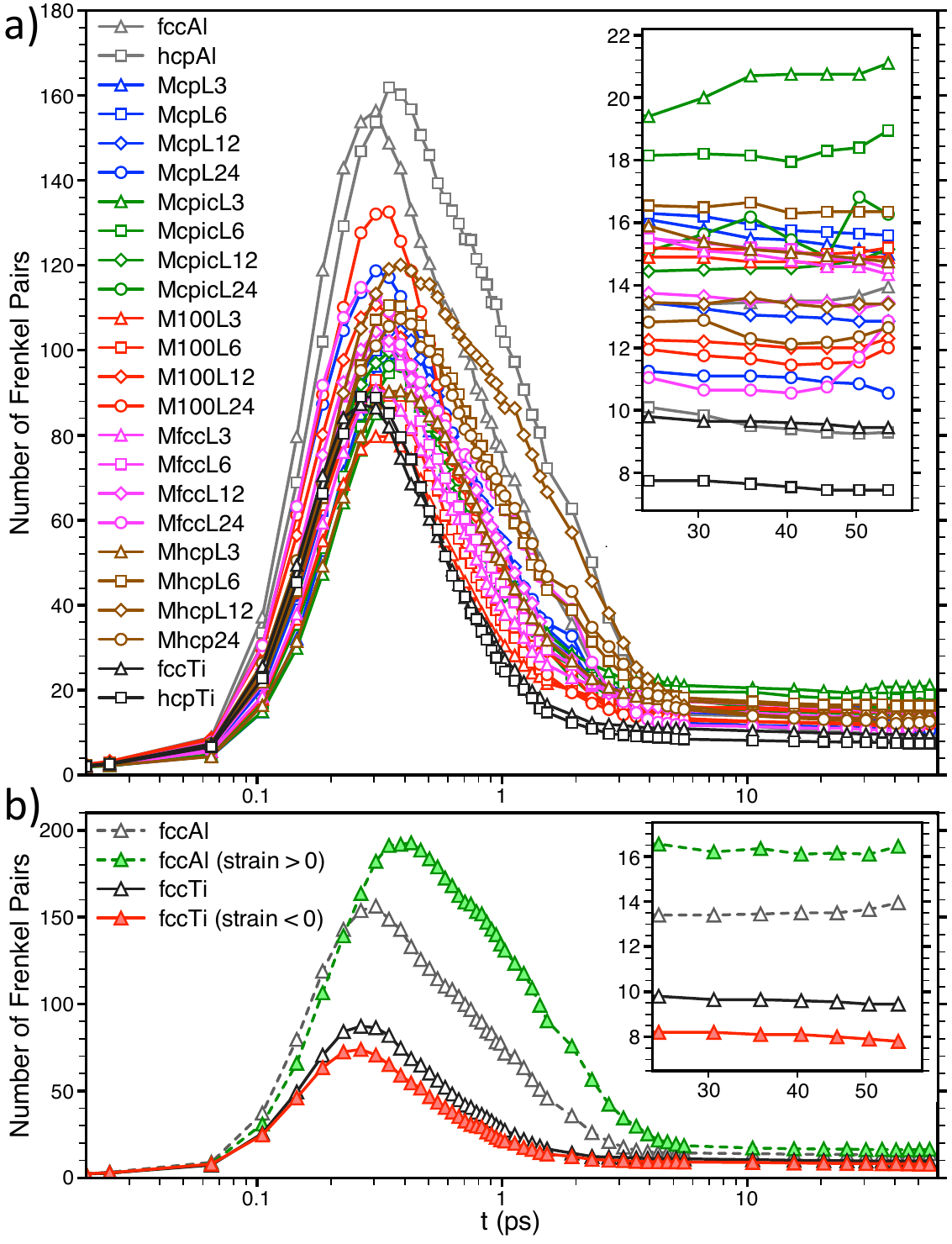}
\caption{\label{figfrenkelpairs} (color online) a) Evolution of the damage production initiated with a 1.5 keV primary-knock-on atom. b) Effect of isotropic strain on defect production in fcc Al ($\epsilon_{iso} = 2.11\%$) and in fcc Ti ($\epsilon_{iso} = -2.07\%$).}
\end{figure}

\begin{figure}
\includegraphics[scale=0.75]{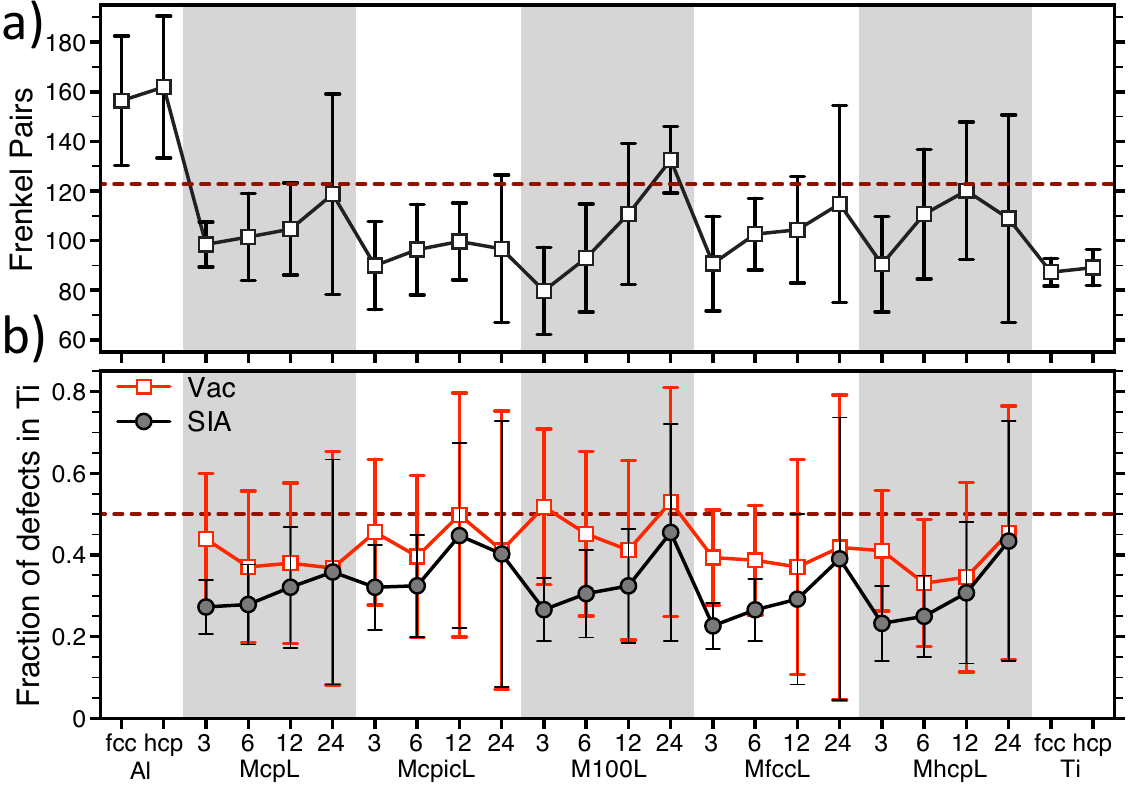}
\caption{\label{fignmax} (color online) a) Number of Frenkel pairs and b) fraction of vacancies and interstitials in the Ti films at maximum damage production near 0.3 ps. The dashed line in a) marks the average between the value of fccAl and hcpTi.}
\end{figure}

All of the displacement cascade simulations were initiated with a 1.5 keV PKA, either Al or Ti. The evolution of damage production (the number of Frenkel pairs) from the simulations is plotted in Figure \ref{figfrenkelpairs}a. The plotted quantities are the average values from the 20 cascade simulations. Different colors represent different systems. In each multilayer system, different film thicknesses are plotted with a different symbol, namely L3 (triangle), L6 (square), L12 (diamond) and L24 (circle). The number of produced Frenkel pairs rises quickly within sub-pico second timespans and reaches maximum $N_{max}$ at approximately 0.3 ps.  Following this stage, most of the displaced atoms quickly recover to lattice sites within several pico seconds. Ti (black curve) exhibits the fastest recovery rate, followed by Mfcc, (Mcpic, Mcp, M100), Mhcp, and finally Al. Figure \ref{fignmax}a shows the $N_{max}$ for all the systems. It appears that the recovery rate is correlated with $N_{max}$, i.e. the rate increases as $N_{max}$ increases. Since one may think of $N_{max}$ as a measure of the size of the damage region, the correlation may be simply a consequence of a fact that thermal recovery takes longer for atoms for larger damage volumes.

To study the effect of strain on the damage production, simulations on fcc Al and fcc Ti with reverse lattice constants were performed. At 300 K, the lattice constant of fcc Al is 4.069 \AA\ while fcc Ti is 4.155 \AA. Systems with a reverse lattice constant: fcc Al with 4.155 \AA\ and fcc Ti with 4.069 \AA\ correspond to isotropically strained systems with $\epsilon_{iso} = 2.11\%$ for Al and $\epsilon_{iso} = -2.07\%$ for Ti. Figure \ref{figfrenkelpairs}b shows the effect of strain on the defect production and recovery rate. The given tensile strain on fcc Al increases $N_{max}$ by (192.8-156.4)/156.4 = 23\% and reduces the recovery rate. On the other hand, the compressive strain on fcc Ti decreases $N_{max}$ by (87.6-73.9)/87.6 = 16\% and increases the recovery rate.

It is worth noting that we have performed simulations to investigate the effect of the PKA direction on the damage evolution curve. Simulations on fcc Al with a PKA initially along $[111]$ compared to $[100]$ yield remarkably similar curves. Tests on M100L6 and McpL6 with PKA direction normal vs tangential to the stacking also produced very similar damage behaviors. This indicates that 1.5 keV used in this study was sufficient to smear out any orientation effect that would otherwise be significant for energies close to the displacement threshold energy $E_t$ (the minimum kinetic energy required to create at least one stable Frenkel pair). Note that $E_t$ varies with the crystallographic direction. We have determined that the average $E_t$ are 20.9 eV (fcc Al), 20.5 eV (hcp Al), 35.4 eV (fcc Ti) and 33.3 eV (hcp Ti).

Figure \ref{fignmax}a shows the maximum number of Frenkel pairs during the cascades near 0.3 ps. The error bars represent the standard deviation from the 20 runs. During this period, a trend of maximum damage production as a function of film thickness was observed within each multilayer system, i.e. $N_{max}$ decreases as the film thickness in each multilayer decreases. This characteristic is particularly pronounced in M100. Note that the value for L24 of Mcpic and Mhcp is smaller than that for the corresponding L12, however the observed trend was still valid within the standard deviation. Within the error bars, $N_{max}$ in all multilayers falls in between that of bulk Al (at the higher end of $N_{max,fccAl} \sim$ 156 pairs) and bulk Ti (at the lower end of $N_{max,hcpTi} \sim$ 89 pairs). What is interesting is that $N_{max}$ in the multilayers is less than the average bulk value $N_{max,bulk} = (N_{max,fccAl} + N_{max,hcpTi})/2$ (dashed line in Figure \ref{fignmax}a). The difference between $N_{max}$ and $N_{max,bulk}$ diminishes as $L$ increases, as expected. Hence the first finding is that the number of displaced atoms during maximum damage production is suppressed in the nanolayered systems and the effect is amplified as the constituent films are made thinner. This finding is generally consistent with both experimental observations \cite{Fu2013PhilMag}, although the research reported here appears to be the first comprehensive simulation study to demonstrate this for point defects (no He atoms).

\begin{figure}
\includegraphics[scale=0.45]{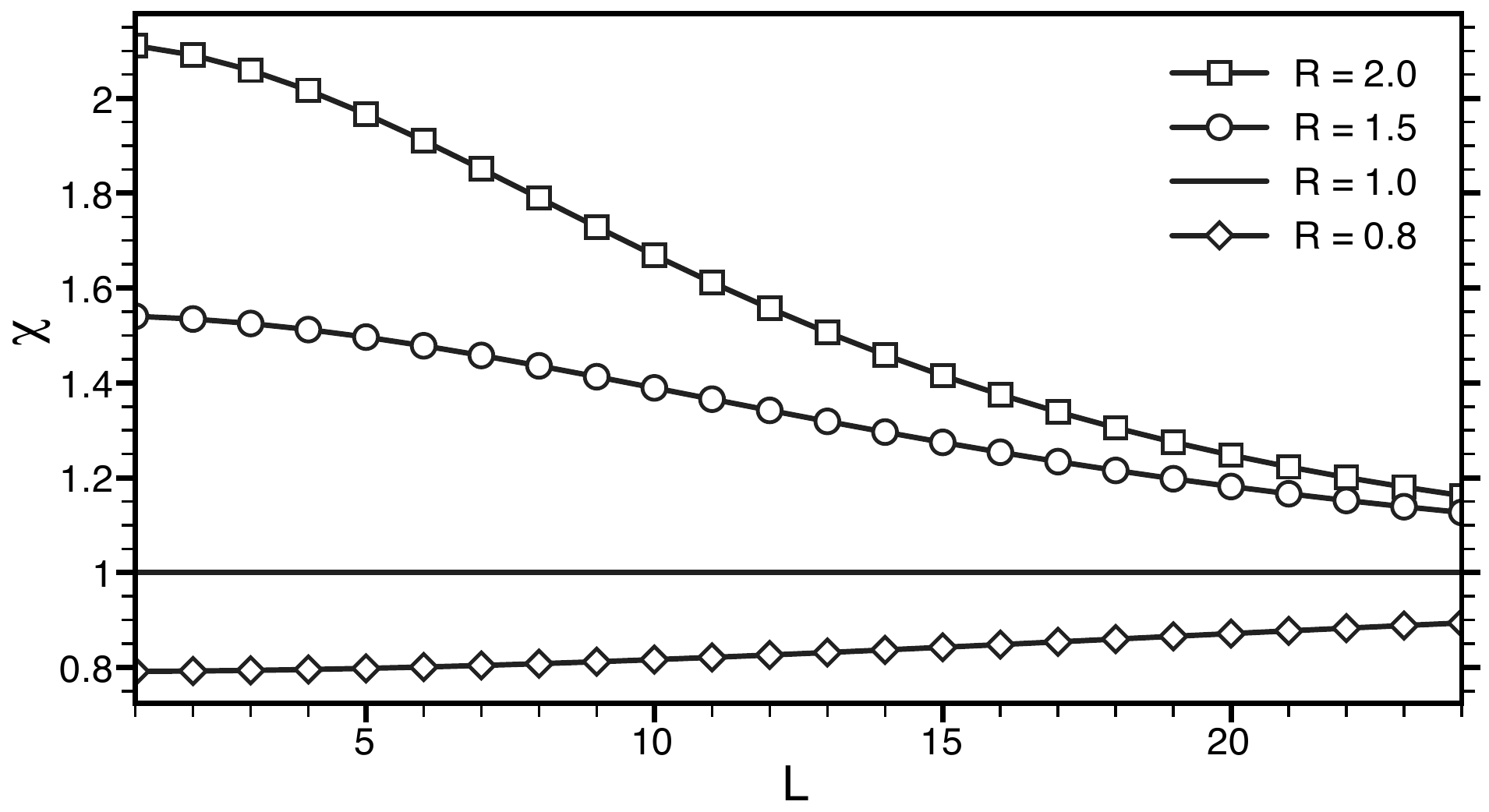}
\caption{\label{figchi} Ratio between energy deposited in the Ti film and in the Al film ($\chi$) as a function of the number of layers per film $L$ for the case of $f_{Al}=0.1$ for several $R = f_{Ti}/f_{Al}$ where $f$ denotes the fraction of energy deposited per layer.}
\end{figure}

To understand the trend of $N_{max}$ as a function of $L$, we constructed a phenomenological model based on the absorption coefficient of a film in slowing down the energetic atoms. Due to its higher $E_t$, it is logical to assume that Ti has a larger absorption coefficient than Al. The reduction of $N_{max}$ as $L$ is decreased may be a consequence of an increased fraction of PKA energy being absorbed in the Ti than in the Al films as $L$ is reduced. Increasing the portion of deposited energy in the Ti films would increase the effective $E_t$ in the multilayer and consequently suppress damage production. With a hypothesis that changing the number of partitions (film interfaces) in the multilayer alters the fraction of energy deposited in the Ti film, the phenomenological model was developed as the following: Let $f_{Ti}$ and $f_{Al}$ be the fraction of energy loss due to (absorbed by) a Ti and Al layer respectively. If we start with a Ti PKA in the first Ti layer and the collision proceeds towards the second Ti layer and so forth, the fraction of energy deposited in the first and second Ti layer is respectively
\begin{eqnarray}
\chi_{Ti,1} &=& f_{Ti}\\
\chi_{Ti,2} &=& f_{Ti}(1 - f_{Ti}),
\end{eqnarray}
and the total fraction deposited in the Ti and Al film in the first pair of Ti-Al film containing $L$ layers each is respectively
\begin{eqnarray}
\chi_{Ti}^{p=1} & = & \sum_{i=1}^L f_{Ti}(1 - f_{Ti})^{i-1} \\
\chi_{Al}^{p=1} & = & \sum_{i=1}^L f_{Al}(1 - f_{Al})^{i-1}(1-f_{Ti})^L
\end{eqnarray}
Similarly, if we start with an Al PKA, the loss fraction in the first $L$-layer Ti and $L$-layer Al is
\begin{eqnarray}
\chi_{Ti}^{p=1} & = & \sum_{i=1}^L f_{Ti}(1 - f_{Ti})^{i-1}(1-f_{Al})^L \\
\chi_{Al}^{p=1} & = & \sum_{i=1}^L f_{Al}(1 - f_{Al})^{i-1}
\end{eqnarray}
Hence, the average loss fraction in the Ti film in the $p$-th pair due to Ti PKA and Al PKA is
\begin{eqnarray}
\chi_{Ti}^{p} & = & \frac{1}{2}[(1-f_{Ti})(1-f_{Al})]^{L(p-1)} \nonumber \\
& & \times \sum_{i=1}^L f_{Ti}(1 - f_{Ti})^{i-1}(1+(1-f_{Al})^L) 
\end{eqnarray}
and finally, the ratio of the energy deposited in the Ti film relative to that in Al from all 48 layers as a function of film thickness is
\begin{eqnarray}
\chi(L) = \sum_{p=1}^{24/L} \chi_{Ti}^{p} \left/  \sum_{p=1}^{24/L} \chi_{Al}^{p} \right.
\end{eqnarray}
Figure \ref{figchi} shows $\chi$ as a function of $L$ for the case of $f_{Al}=0.1$ for several $R = f_{Ti}/f_{Al}$ ratios: 0.8, 1.0, 1.5 and 2.0. All $\chi$ curves approach unity in the limit of infinite $L$. For $f_{Ti}>f_{Al}$ case, it can be seen that $\chi$ increases as $L$ is decreased confirming the hypothesis that an increased fraction of the PKA energy is deposited in the Ti films as the film thickness is reduced. 

We realize that in multilayers, the absorption coefficient for each layer varies even within the same film due to the different atomic environments  and strain fields experienced by each layer. Moreover, determining the value of $f_{Ti}$ and $f_{Al}$ of each layer is not straightforward. However, we believe that the underlying physics captured in the model sufficiently describes the observed trend of $N_{max}$ vs $L$, at least qualitatively. In other words, in a multilayer system, even though the proportion of the constituent materials is kept the same, the response of the system can be driven closer to the more dominant materials by reducing the thickness of each film. In this case, $N_{max}$ in L3 is the closest to that in pure bulk Ti.

\begin{figure}
\includegraphics[scale=0.75]{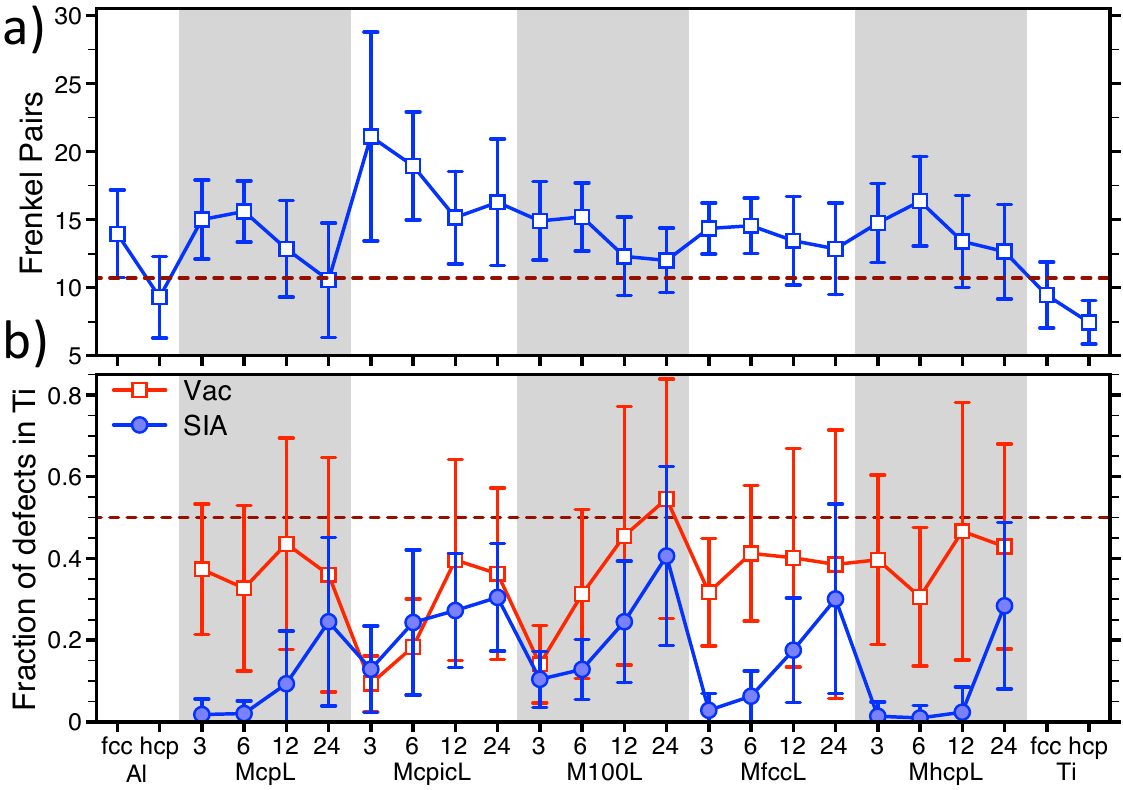}
\caption{\label{figpartitionend} (color online) a) Number of Frenkel pairs and b) fraction of vacancies and interstitials in the Ti films at the end of simulation after 55.5 ps. The dashed line in a) marks the average between the value of fccAl and hcpTi.}
\end{figure}

The number of surviving defects at the end of simulations ($N_{end}$) is plotted in Figure \ref{figpartitionend}a. The values for the bulk structures are $N_{end,fccAl} \sim 14$ and $N_{end,hcpTi} \sim 7$. The dashed line at 10.5 marks the average bulk value $N_{end,bulk} = (N_{end,fccAl} + N_{end,hcpTi})/2$. In all of the multilayers, even though $N_{max} < N_{max,bulk}$, the surviving number of defects is larger than $N_{end,bulk}$. This indicates that the vacancy-interstitial recombination in these multilayers is inhibited. The defect spatial distribution, defect cluster morphology as well as the strain field may contribute to altering the defect recombination process.

To better understand how defects are distributed in the multilayer, we present an analysis of defect distribution at the maximum damage production regime as well as at the end of the simulations. The defect distribution near the maximum damage production is presented in Figure \ref{fignmax}b. The plotted quantities are the fractions of vacancies (square marks) and self-interstitial atoms SIAs (circles) in the Ti films. The plot shows that there are fewer vacancies in Ti films than in Al films. This is understood from the larger $E_t$ of Ti. There are also fewer interstitials in the Ti films than in Al films. In fact, in the Ti films the number of interstitials is even smaller than the number of vacancies. This indicates that there is an imbalance flux of displaced atoms from Ti films to Al films. The degree of imbalance systematically increases as the film thickness decreases (as the number of interfaces is increased). Near the interface, we observe that displacing a Ti atom from a Ti film to the Al region is energetically favorable compared to the opposite process. This was caused by the fact that it is easier for heavier Ti atoms to displace lighter Al atoms whose $E_t$ is also smaller than otherwise. Hence, the interface has induced a preferential drift of SIAs from the Ti to the Al films causing imbalance population of SIAs relative to vacancies in a particular film. We refer this phenomenon as "partitioning" effect. It is expected then that the partitioning effect inhibits defect recombination in multilayers relative to multilayer systems that do not experience such defect partitioning. However, based on the varied and generic observations by many others on the asymmetry of damage in many of the multilayered systems studied to date, both miscible and immiscible, we expect that this is an important result that emerged from this study that has wide-ranging implications for multilayer radiation damage tolerance. In general, defect partitioning contributes to multilayer degradation and failure since it directly leads to damaging differential response.

Figure \ref{figpartitionend}b shows the vacancy-SIA fraction imbalance in the Ti films at the end of the simulations. Unlike the imbalance curve near the maximum damage production (Fig. \ref{fignmax}b), the imbalance at the end of simulation shows two different characteristics depending on the multilayer system. Firstly, in Mcp, Mfcc and Mhcp, the imbalance is still evident, in fact it is more pronounced due to the much smaller fraction of SIAs that survives in the Ti films. As the result of the partitioning effect, in the Ti films the number of vacancies is more than what is needed for the recombination, while in the Al film there are more SIAs than the available vacancies to recombine. The second characteristic of the imbalance curve is observed in Mcpic and M100. In the M100, even though the fraction of vacancies in the Ti films is still larger than the fraction of SIAs, the difference diminishes towards L3. In the Mcpic, the fraction of vacancies in the Ti films becomes comparable to that of SIAs. In this case, it appears that a portion of SIAs in the Al films recombine with vacancies in the Ti films, particularly those at the interface, mostly during the early stages of recovery. The different characteristic of fraction imbalance at the end of simulation between Mcp-Mfcc-Mhcp and M100-Mcpip may be related to the strain in the film. From Table \ref{tablestrain}, in the first group of multilayers, Ti films are compressed in both basal directions ($\epsilon_x^{Ti} = \epsilon_y^{Ti} < 0$) and the compressive strain increases as film thickness decreases. The opposite case occurs in the second group of multilayers: in the M100 $\epsilon_x^{Ti}$ gradually becomes $>$ 0 at L3, while in Mcpic even though $\epsilon_1^{Ti}$ is slightly $<$ 0, $\epsilon_2^{Ti}$ is $>$ 0. In addition, unlike in all other systems in which Al films are under tension, Al film in the Mcpic is slightly compressed. We believe that the reduction of the exclusivity of compressive strain in the Ti films (on one hand) and tensile strain in the Al films (on the other hand) in the M100 and Mcpic multilayers plays a role in reducing the vacancy-SIA fraction imbalance by allowing a portion of the SIAs in the Al film to recombine with vacancies in the Ti film near the interface during the recovery process.

\begin{figure}
\includegraphics[scale=0.32]{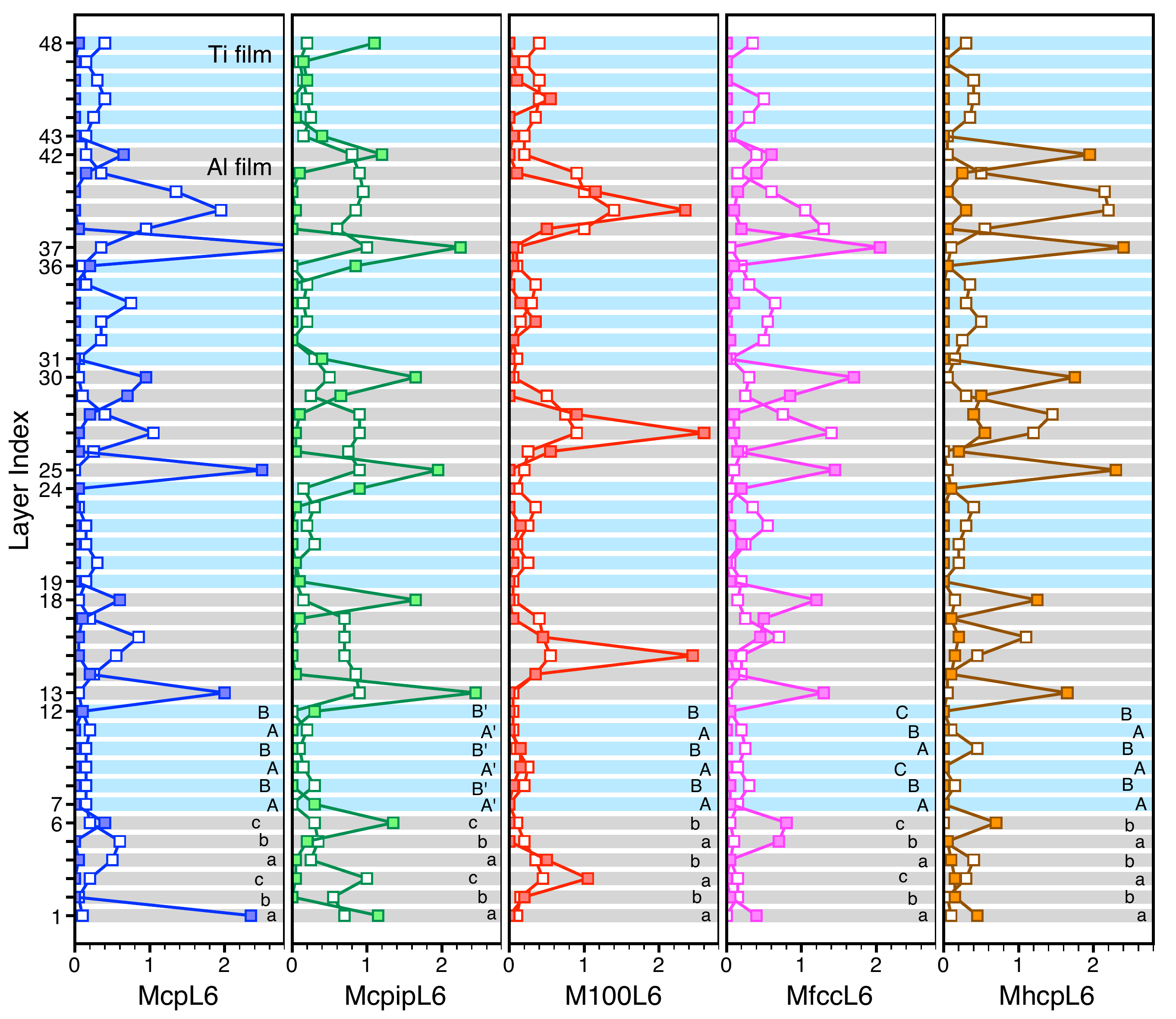}
\caption{\label{figbinzL6} (color online) Distribution of vacancies (hollow squares) and interstitials (filled squares) at the end of simulation after 55.5 ps.}
\end{figure}

Besides the differential defect distribution (partitioning effect), the different strain levels that are experienced by each layer in the film can significantly affect defect migration. The SIAs may either preferentially migrate to the interface or to the middle of the film away from the interface. To study defect migration, the number of surviving vacancies and interstitials at the end of the simulations in each layer along the multilayer stacking direction is calculated. Figure \ref{figbinzL6} shows the result for L6 in each multilayer (other film thicknesses show similar distributions). In Figure \ref{figbinzL6} the vacancies are plotted with hollow marks while interstitials are presented as filled marks. In all systems, the multilayer starts with Al film at the bottom (gray) followed by Ti film (blue). The stacking sequence is included in the plot for clarity. As has been discussed, the majority of the defects are found in the Al films. Figure \ref{figbinzL6} also reveals that in all mutilayers except the M100, the SIAs are preferentially found at the interface layer in the Al film. Meanwhile, for M100, the interstitials preferentially migrate to the middle of the Al films. The interstitials in Ti film in M100 also migrate to the middle of the Ti film even though the number is much smaller than in Al films.

To understand why the interstitials in the Al films migrate to the middle of the film in M100 while they migrate to the interface layer in all the close-packed multilayers, formation energies of dumbbells in McpL6 (to represent the multilayers of close-packed layers) and M100L6 were calculated. A single interstitial was added to the system and the atoms were relaxed via energy minimization. The results are presented in Table \ref{tabledb}.

\begin{table}
\caption{\label{tabledb}
Dumbbell orientation and formation energy $E_f$ in McpL6 and M100L6 multilayer. Layer indexing in the stacking starts from bottom to top: Al-1$\rightarrow$Al-6, Ti-7$\rightarrow$Ti-12 and so on. The * indicates that a dumbbell stabilizes in a different layer than its initial position during relaxation. All Miller indices are with respect to a cubic system.}
\begin{ruledtabular}
\begin{tabular}{cccc}
Layer &dumbbell&$E_{f}$ (eV)&bond (\AA)\\
\hline
McpL6&&&\\
Al-1& [11$\bar{2}$] (Al-Al)& 1.27&2.27\\
Al-2& [11$\bar{2}$] (Al-Al)& 1.68&2.26\\
Al-3& [100] (Al-Al)             & 1.82&2.35\\
Al-4& [100] (Al-Al)             & 1.84&2.35\\
Al-5& [11$\bar{2}$] (Al-Al)& 1.82&2.26\\
Al-6& [10$\bar{1}$] (Al-Al)& 1.64&2.27\\
Ti-7*& [10$\bar{1}$] (Al-Al) in Al-6& 1.07&2.27\\
Ti-8& [11$\bar{2}$] (Ti-Ti)           & 3.31&2.26\\
Ti-9& [11$\bar{2}$] (Ti-Ti)           & 3.28&2.32\\
Ti-10& [11$\bar{2}$] (Ti-Ti)        & 3.27&2.26\\
Ti-11*& [11$\bar{2}$] (Al-Al) in Al-13& 0.59&2.26\\
Ti-12*& [11$\bar{2}$] (Al-Al) in Al-13& 0.59&2.26\\
\hline
M100L6&&&\\
Al-1*& [001] (Al-Al) in Al-2& 2.03&2.29\\
Al-2*& [001] (Al-Al) in Al-3& 1.84&2.34\\
Al-3& [100] (Al-Al)           & 1.78&2.36\\
Ti-7*& [001] (Ti-Ti) in Ti-8& 3.04&2.32\\
Ti-8*& [001] (Ti-Ti) in Ti-9& 2.91&2.33\\
Ti-9& [100] (Ti-Ti)           & 2.88&2.34\\
\end{tabular}
\end{ruledtabular}
\end{table}

%
%
%

\begin{figure}
\includegraphics[scale=0.55]{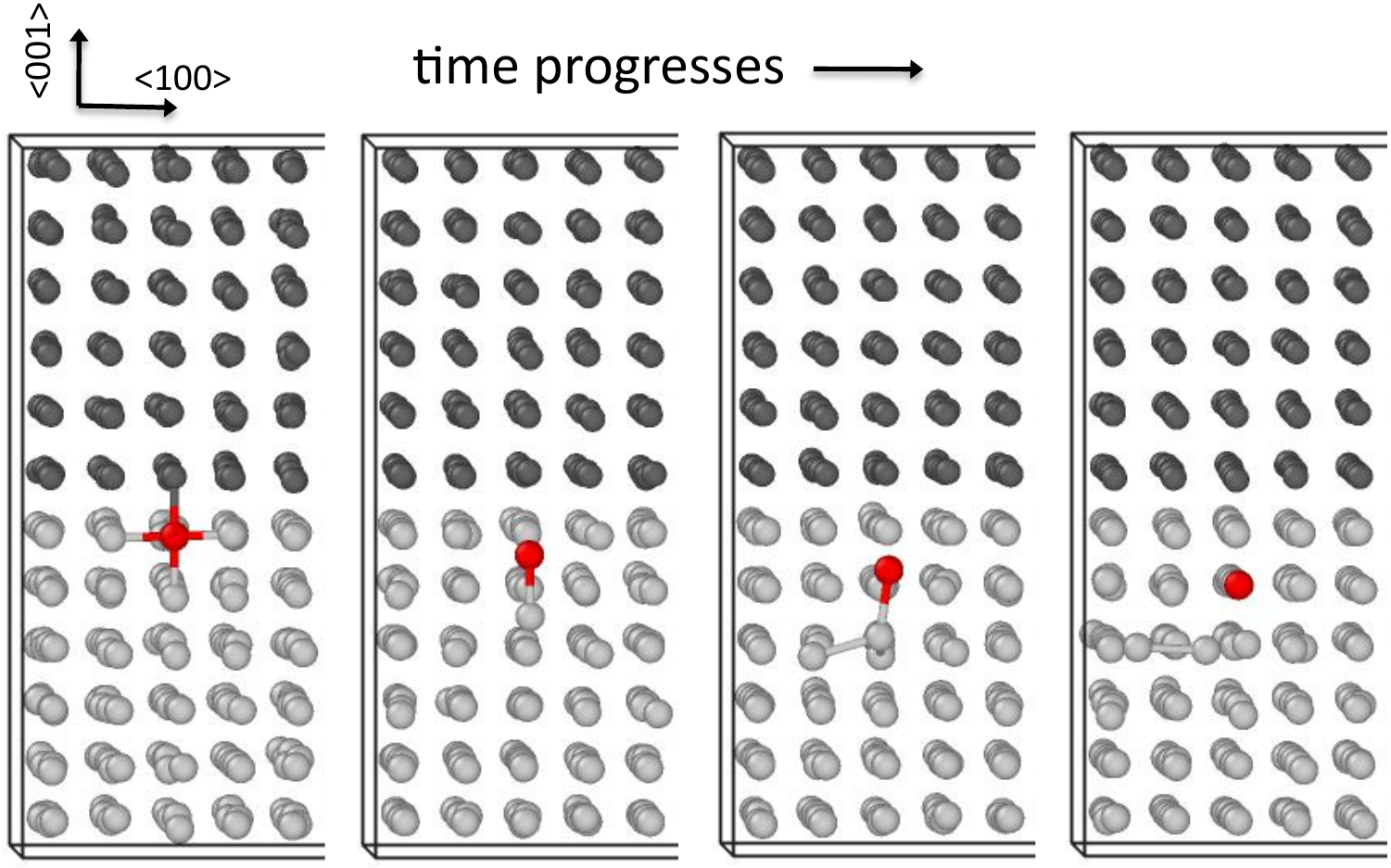}
\caption{\label{figdbM100} (color online) Migration pathway of an Al atom (red) initially at octahedral interstitial site in M100L6 resulting in the formation of a [100] dumbbell in the middle layer of Al film (light gray).}
\end{figure}

In McpL6, the preferred location for the Al-Al dumbbell is at the interface layer of Al films with [11$\bar{2}$] orientation (in plane with the close-packed layer) with a formation energy of 1.27 eV. Note that Miller indices used to describe the dumbbell orientation are with respect to a cubic system.  In the middle of the Al films, the preferred Al-Al dumbbell orientation is [100] with a formation energy of 1.84 eV ($\sim$0.6 eV higher than that at the interface layer). The situation in the Ti films is even more pronounced, if a Ti interstitial is found in the middle layer of a Ti film, it forms a [11$\bar{2}$] dumbbell with  formation energy of 3.28 eV. If the Ti interstitial is placed in the Ti layers 11 or 12 (interface layer), it initiates a sequence of relaxations so that one Ti atom occupies a lattice site in the Al films leaving an Al interstitial in Al layer 13 (interface layer) forming an Al-Al [11$\bar{2}$] dumbbell with formation energy of only 0.59 eV. The opposite trend of dumbbell formation energy is found in M100L6. In this case, the Al-Al dumbbells are most stable in the middle of Al films forming in [100] orientation with formation energy of 1.78 eV compared to 2.03 eV for [001] dumbbell found at the interface layer. Within the Ti films, the middle layer also provide the stable location for Ti-Ti dumbbell forming in [100] orientation with formation energy of 2.88 eV compared to 3.04 eV for [001] dumbbell found at the interface. Hence, it is clear why in the close-packed multilayers, the SIAs migrate to the interface while in the M100 system they migrate away from the interface to the middle of the film. Figure \ref{figdbM100} illustrates the migration process of an Al interstitial initially placed in the Al interface layer (red atom) that results in the formation of a [100] dumbbell in Al middle layer (gray).

\subsection{Experimental Results}
\begin{figure}[htb]
\includegraphics[scale=0.68]{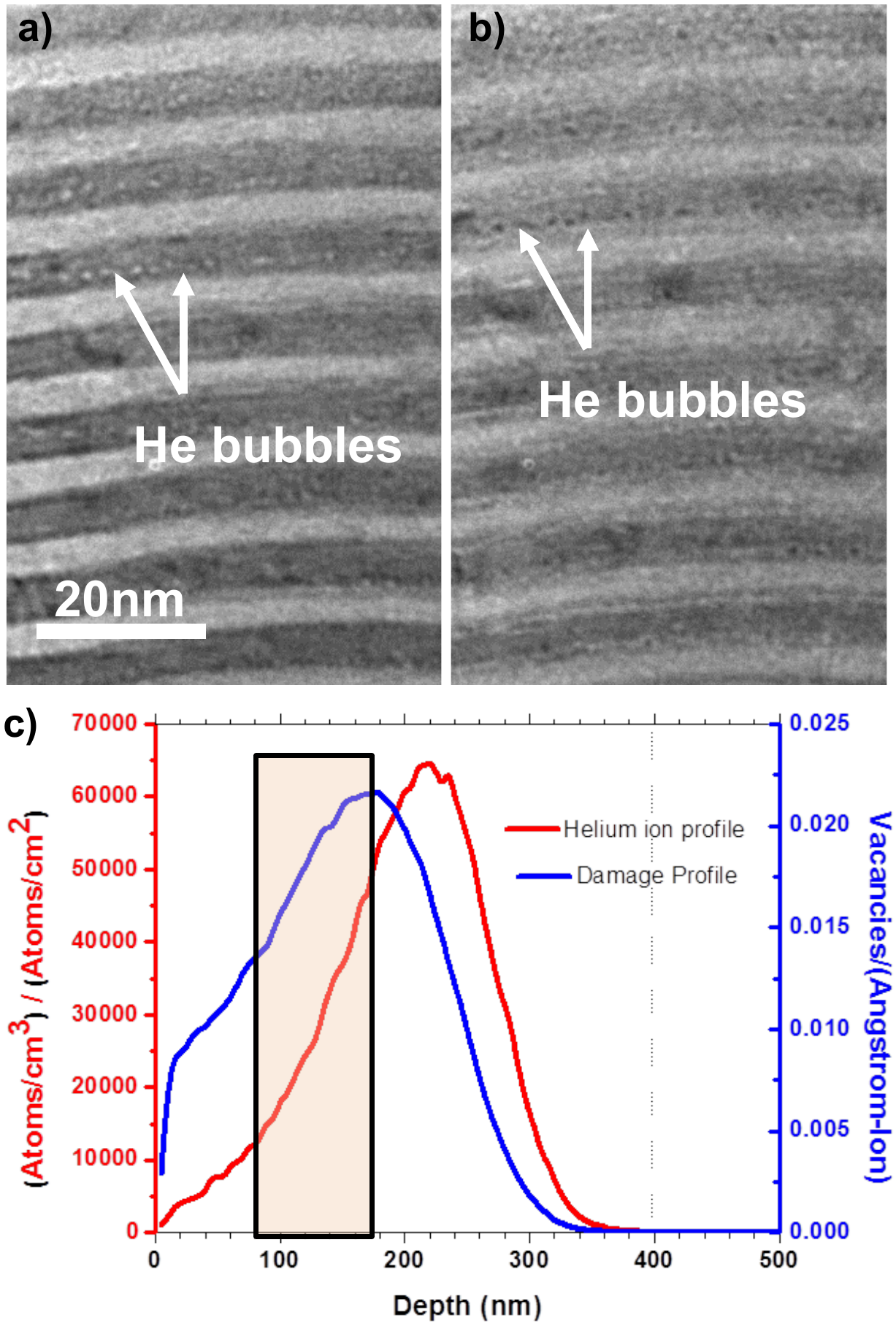}
\caption{\label{figHe} (color online) a) Underfocused and b) overfocused TEM images showing the location of Helium bubbles in the upper middle region of the Ti/Al multilayer sample 5 nm per film. In the underfocused image the He bubbles appear bright and in overfocused image He bubbles have dark contrast. The bubbles are preferentially segregated to the darker contrast Ti layers. c) SRIM simulation showing He ion profile and damage profile.}
\end{figure}

Figure \ref{figHe}a shows the cross-section image of Al-Ti multilayer sample with thickness of 5 nm per film ($\sim$21 layers) obtained with underfocused TEM. The image was taken after He irradiation with dose 10$^{16}$ atoms/cm$^2$ at room temperature. In this image, Ti films appear darker than Al due to atomic number contrast. In the Ti films, bright spots can be seen that represent He bubbles. The diameter of the bubbles is $\sim$1 nm. This result is intriguing for a reason that due to a lower displacement threshold energy of Al compared to Ti, the nucleation of small bubbles via a kickout mechanism (a cluster of He atoms displacing a host atom from its lattice site) would be expected to occur in Al films. As a reference, to create a 1-nm bubble, $\sim$56 Ti atoms or $\sim$59 Al atoms would need to be displaced. 

The fact that the bubbles are found in the Ti films suggests that the distribution of He atoms during the irradiation plays a major role in determining the morphology and location of the bubbles. It is possible that the larger stopping power of Ti films has caused the He atoms to be stopped and contained in the Ti films more effectively than in Al films. In this scenario, the necessary space needed for the bubbles is created not via a kickout mechanism but rather during the collision cascade itself. In this stopping process, the impinged Ti atoms may remain in the Ti films or be displaced to the Al films. If the impinged Ti atoms can be displaced to the Al films, this process will greatly favor the creation of the necessary excess volume for the He atoms to form small bubbles in Ti films. The defect imbalance that is observed in the simulations provides a clear proof that displacing Ti atoms from Ti films to Al films is indeed easier than the reverse. 

In Figure  \ref{figHe}a, the He bubbles are arranged in a row with a somewhat regular spacing between the bubbles. More importantly, the bubbles are located near the interface towards the Al films below the Ti films where they reside, i.e. the location of the bubbles is biased towards the direction of the irradiation. This provides another clue that He bubble formation in the Ti films is associated with a preferential flux of SIAs from Ti films to Al films during the irradiation as described above. 

Figure \ref{figHe}c shows the He ion profile and the damage profile obtained with SRIM simulations. The rectangular block represents the region of Al-Ti sample that was imaged. The damage profile corresponds to the distribution of the vacancies.

\section{Conclusion}
This work has revealed several important properties of Al/Ti nanolayers in response to radiation damage, the most important finding being the observation of strong defect partitioning during collision cascades that imparts a strong asymmetry in radiation damage. During maximum damage production initiated with 1.5-keV PKA, asymmetry in the point defects creation between the two dissimilar materials Al or Ti is observed with $\sim$60\% of vacancies are created in Al films while $\sim$70\% of interstitials are created in Al films. The excess interstitials in the Al films is a direct consequence of the preferential flux of displaced atoms in this nanolayered system. He irradiation experiments at a low dose of 10$^{16}$ atoms/cm$^2$ and at room temperature were performed to investigate the interface effects on radiation damage. The experimental data shows a formation of $\sim$1-nm diameter bubbles in the Ti films near the interface. The results from the simulations and experiments seem to suggest that the He bubble formation is associated with the preferential flux of SIAs during the irradiation in that the He atoms impinge on Ti films and displace Ti atoms into the Al films. This is further supported by the location of the bubbles being near the interface and biased towards the direction of the irradiation. 

In all of the interface models in this study, the number of Frenkel pairs created during maximum damage production is smaller than the bulk average of the constituent materials. This difference is amplified for thinner films in accordance with experimental observations on other multilayer systems. This observation is understood using a phenomenological model that Ti exhibits a larger stopping power than Al and that the fraction of energy deposited in Ti films increases as the films are made thinner. On the other hand, the number of surviving Frenkel pairs at the end of simulations is larger than the bulk average. The difficulty for anti-defect recombinations is caused by defect partitioning in which in Al films there are too many interstitials than the available vacancies while the opposite applies in Ti films. This defect partitioning increases with the increasing number of interfaces (thinner films) resulting in more than 90\% of surviving interstitials located in the Al films for nanolayers with $\le$ 6 layers per film.

These simulation results, when considering all the other experimental and modeling results for nanolayered systems, suggest that, in addition to interface structure and chemical mixing, we add degree of defect partitioning to the list of desirable system properties in the design of radiation tolerant material systems. Differential material responses that grow with increased radiation dose are not a recipe for stable damage tolerant material systems.

\begin{acknowledgments}
This research was supported by the award BRCALL08-Per4-E-1-0062 from Defense Threat Reduction Agency (DTRA). A portion of this research was performed using Olympus supercomputer at EMSL ($\#$44724), a national scientific user facility sponsored by the Department of Energy's Office of Biological and Environmental Research and located at Pacific Northwest National Laboratory. We thank Dr. Tamas Varga for fruitful discussions.
\end{acknowledgments}

%
\end{document}